\documentclass[12pt,preprint]{aastex}
\shorttitle{Lyman $\beta$ fluorescence in Be stars}
\shortauthors{Mathew et al.}

\begin{document}

\title{A study of the role of Lyman $\beta$ fluorescence on O{\sc i} line strengths in Be stars}

\author{Blesson Mathew$^1$, D. P. K. Banerjee$^1$, A. Subramaniam$^2$ and N. M. Ashok$^1$}

\affil{$^1$Astronomy and Astrophysics Division, Physical Research Laboratory, 
Navrangapura, Ahmedabad - 380 009, Gujarat, India}
\affil{$^2$Indian Institute of Astrophysics, Bangalore - 560 034, India}
\email{blesson@prl.res.in}

\begin{abstract}
The possibility of the Ly$\beta$ fluorescence
mechanism being operational in classical Be stars and thereby contributing to the
strength of the O{\sc i} 8446 \AA~line has been recognized for long.
However this supposition needs to be quantified by 
comparing observed and predicted O{\sc i} line ratios. In the present work,
optical and near-infrared spectra of classical Be stars are presented. 
We analyse the  observed strengths of the O{\sc i} 7774, 8446,
11287 and 13165 \AA~lines which have been theoretically proposed as 
diagnostics for identifying the excitation mechanism. We have considered and examined 
the effects of Ly$\beta$ fluorescence, collisional excitation, recombination 
and continuum fluorescence on these O{\sc i} line strengths. 
From our analysis it appears that the
Ly$\beta$ fluorescence process is indeed operative in Be stars.
\end{abstract}


\section{Introduction}
Classical Be (CBe) stars are non-supergiant B-type stars whose spectrum, to 
enable a definition in a broad sense, show or have shown Balmer lines in 
emission at some stage \citep{collins87}. These and other emission lines 
are formed in a circumstellar gaseous decretion disk which is also the source for a significant
infrared continuum excess arising from free-free and bound-free emission. 
It has been considered for long that the disk is formed as the outcome of Be
stars being fast rotators. However, whether formation of a geometrically thin disk can be caused solely 
by rotation is not as yet a resolved issue (for e.g., see the review by \citet{porter03}).
Recent studies suggest that CBe stars are rotating close to their critical 
velocities and other mechanisms like non-radial pulsations, magnetic fields or
binarity can possibly contribute to explaining the Be phenomenon \citep{meilland12}.

The dynamics and nature of the circumstellar disk can be understood in a
better manner from spectroscopic observations. One of the initial efforts in this
direction is by \citet{slettebak82} who estimated the spectral types and
rotational velocities of 183 Oe, Be, A-F type shell stars brighter than 6
magnitude. Subsequently, several other detailed spectroscopic surveys have
been done both in the optical (for example \citet{Andrillat82},
\citet{Andrillat88}, \citet{Dachs86}, \citet{Dachs92}, \citet{Banerjee00},
\citet{hanuschik86}, \citet{hanuschik87}) and in the
near-infrared (for example \citet{clark00}, \citet{steele01}) to understand 
the behavior and temporal evolution of different lines and in general to
understand the Be phenomenon. Our present study focuses on studying the 
strengths of certain O{\sc i} lines that are commonly seen in the optical and
near-infrared spectra of CBe stars viz. the O{\sc i} 7774, 8446, 11287 and
13165 \AA~lines. The motivation for studying these lines is discussed below in
greater detail. Our optical sample of CBe stars is drawn from earlier studies
of \citet{mathew08} and \citet{mathew11} 
wherein we had surveyed, identified and obtained spectra of 150 CBe stars in selected open clusters.

Among the O{\sc i} lines one may begin by considering  O{\sc i} $\lambda$8446 
which is present in emission in the spectra of several astrophysical sources 
viz. emission-line stars, H{\sc ii} regions, planetary nebulae,
novae, Seyfert galaxies and QSOs. It is an interesting line because
there has been considerable discussion in the literature on the possible mechanisms which excite it.
It is generally accepted that
Ly$\beta$ fluorescence, apart from other mechanisms like collisional 
excitation, recombination and continuum
fluorescence, can greatly contribute to the observed strength of the line.
The Ly$\beta$ fluorescence mechanism was proposed by \citet{bowen47}
wherein due to the near coincidence of wavelengths, Hydrogen Ly$\beta$
photons at 1025.72 \AA~can pump the O{\sc i} ground state resonance line at
1025.77 \AA~thereby populating the O{\sc i} 3d$^3$D$^0$ level.
The subsequent downward cascade produces the 11287, 8446 and 1304 \AA~lines in
emission. This was referred to as a PAR process
(photoexcitation by accidental resonance) by \citet{bhatia95} (hereafter BK95)
who constructed a detailed model for neutral oxygen and computed expected
strengths of the various O{\sc i} lines when  collisional excitation was the
sole excitation mechanism (BK95) and followed it with a second study in which  the PAR
process was also considered (\citet{kastner95}, hereafter KB95). 
In the present study, we have used the results from these two studies 
to understand the neutral O{\sc i}
spectrum of CBe stars and discriminate whether the PAR process is indeed
responsible for the observed strength of the 8446 \AA~line in these stars.
That the Ly$\beta$ fluorescence process could be operating in CBe stars has been
suggested since long. For example as early  as 1951, \citet{slettebak51}
in a survey of 25 CBe stars pointed out the  anomalies in
the strengths of the 7774 \AA~and 8446 \AA~lines with the well known tendency of $\lambda$8446
to go into emission more readily than  $\lambda$7774. This  was attributed to the
Ly$\beta$ process increasing the strength of the 8446 \AA~line. A similar conclusion was
reached by \citet{burbidge52} while studying the CBe star $\chi$ Oph.
However, as of today, observational strengths  of the O{\sc i} lines in Be stars do
not appear to have  been compared in a comprehensive manner with the
detailed theoretical calculations, such as those of BK95 and KB95, to establish whether the
Ly$\beta$ fluorescence process really affects the neutral O{\sc i} spectrum. 
In the present work, we study the behavior of the O{\sc i} 7774, 8446 \AA~lines
in the optical and the O{\sc i} 11287 and 13165 \AA~lines in the near infrared 
and show that there is convincing evidence indicating that the
Ly$\beta$ fluorescence process is indeed operational in CBe stars.

\section{Observations \& Analysis}
Our optical sample of CBe stars is drawn from the study of \citet{mathew08}
who performed a survey to identify CBe stars in 207 open clusters using
slitless spectroscopy. The spectroscopy of 150 CBe stars in 39 open
clusters in the wavelength range 3800 -- 9000 \AA~ is presented in
\citet{mathew11} -- the optical spectra analysed here are from this study. 
Briefly, these spectroscopic observations were done using the
HFOSC imager-spectrograph instrument
available with the 2.0m Himalayan Chandra Telescope (HCT), located at Hanle, India.
The CCD used for imaging is a 2 K $\times$ 4 K CCD,
with a  pixel size of 15 $\mu$ and an image scale of 0.297 arcsec/pixel.
The spectra of CBe stars were taken
at an  effective resolution of 7 \AA~around the H$\alpha$ line. More details on the
observations are presented in \citet{mathew11} and details on the HFOSC are
available at the website \url{www.iiap.res.in}.
All the observed spectra were wavelength calibrated and corrected
for instrument sensitivity using the Image Reduction and Analysis Facility (IRAF) tasks.
IRAF was also used for all further data reduction and analysis including measurements of the 
equivalent widths ($W$) of the spectral lines - the 
typical error in the measurement of $W$ values is around 10\%.

The $J$ band spectra were obtained at a resolution of $\sim$ 1000 using the Near-Infrared
Imager/Spectrometer with a 256$\times$256 HgCdTe NICMOS3 array, mounted on the 1.2m Mt. Abu
telescope. For each object a set of two spectra were taken with the star dithered to two different
positions along the 1 arcsec wide slit. Spectral calibration was
done using OH airglow and telluric lines which register with the stellar
spectrum. Following the standard practice used in the near-IR, the telluric
lines present in the object spectra were removed through a rationing process, in which the 
object spectrum is divided by the standard star spectrum. 
The standard star was always observed at similar 
airmass as the object and prior to rationing the hydrogen lines
in its spectrum were removed by interpolation. The rationed spectra were 
finally multiplied by a blackbody spectrum,
corresponding to the effective temperature of the standard star, to yield the
final spectra. The spectra presented here are for the CBe stars HD 10516, HD 12856, HD 19243, 
HD 35345, $\beta$ Mon A and $\beta$ Mon C which were observed on 
2009 November 8, 2010 December 12, 2009 November 7, 2010 December 13, 
2010 October 22 and 2010 October 22 respectively. 
The corresponding standard stars used were 
SAO 22696, SAO 22859, SAO 12438, SAO 57819, SAO 151911 and SAO 151911. 
The spectral data reduction and analysis were done using IRAF tasks. 
In this work, we have presented the $J$ band spectra of a few field CBe stars which have
been observed as part of an ongoing program to study the $J$ band spectra of 
CBe stars. A detailed analysis of the $J$ band spectra of CBe stars will be
presented later.

\section{Results}
In the optical, the focus was on studying the strengths of the O{\sc i} 7774 and
8446 \AA~lines for reasons outlined below.
To study the effects of the H Ly$\beta$/O{\sc i} PAR process, KB95
have concentrated on five visible/infrared
lines most relevant to the process. These are the forbidden line at 6300 \AA~which serves
as a non-fluorescent standard when the line is available, the allowed
8446 \AA~transition which is an expected fluorescent product, the
allowed multiplet at 7774 \AA~conventionally,
thought to be independent of the H Ly$\beta$/O{\sc i} PAR process, the allowed transition at 11287
\AA~which is the primary cascade line in the PAR process and an additional IR line at
11298 \AA. The 6300 \AA~line is a suitable line of choice for examining
effects of the PAR process in novae
since it is seen in their spectra. However this line is hardly seen in our CBe
spectra except in few and isolated cases where it is extremely weak; it is
hence not a suitable diagnostic line for our purpose.
On the other hand the 7774, 8846 \AA~lines are prominently seen in CBe star
spectra and, as emphasized by KB95, their wavelength proximity introduces
minimal errors when their strengths are compared even if the line intensities
are not deredenned. We have also used the 11287, 13165 \AA~near-IR $J$ band
lines  since the ratio of their line  strengths is an effective discriminator
between Ly$\beta$ and continuum fluorescence as described in  \citet{grandi75}
and \citet{strittmatter77}. We note that a comprehensive $J$ band
spectral study of  Be stars is unavailable though the $H $\&$ K$ bands are
well investigated (\citet{clark00}, \citet{steele01}) as well as
the $L$ band \citep{granada10}. For that matter, the $J$ band spectra of
even isolated CBe stars are not readily encountered in the literature.

In the present study, CBe stars are broadly classified 
into three groups based on the nature of 
O{\sc i} 7774 and 8446 \AA~line profiles. Group I comprises of 83 stars which 
show both the O{\sc i} 7774 and 8446 \AA~lines in emission (Figure 1, for a
few examples). Out of these 83, 6
stars show the O{\sc i} lines very weakly in emission making it difficult to
measure the equivalent widths with any accuracy. 
Therefore we have omitted these stars and considered only 77 stars in Group I
for the present analysis. 
Group II show O{\sc i} 8446 \AA~in emission and 7774 \AA~in absorption 
with 26 CBe stars belonging to this category (top panel
of Figure 2). Group III mostly includes 
objects (23 CBe stars) in which O{\sc i} 8446 \AA~is in emission but
7774 \AA~is not prominent enough for a clear-cut classification, with the
spectra we have, into a
absorption or emission group (bottom panel of Figure 2). 
For comparing strengths of the 7774 and 8446 \AA~lines,
it is meaningful to use only objects where both lines are in emission; we have
thus restricted the analysis described below to objects from the first group.

\subsection{Correction for measured O{\sc i} equivalent widths}
To estimate the equivalent width of O{\sc i} $\lambda$8446, it is
necessary to deblend this line from Paschen 18 (P18, 8437 \AA).
It is possible to estimate the value of $W$(P18) by linearly interpolating between
measured equivalent widths of
the flanking lines P17 (8467 \AA) and P19 (8413 \AA) 
since the emission monotonically decreases along the Paschen series in this
region as indicated by \citet{briot81a} and \citet{briot81b}. 
This value of $W$(P18) may then be subtracted from the combined equivalent width of
$W$(8446 + P18) to get the intrinsic value of $W$(8446) in emission. 
To confirm the validity of this procedure we have plotted in Figure 3 the
ratio of the mean equivalent widths of P14, P17, P19, P20, P21 normalized to the
equivalent width of P17. In this figure we do not include P15 and P16 because
they are blended with the Ca{\sc ii} lines at 8542 and 8498 \AA~. 
The mean equivalent widths of the Paschen lines were computed from a sample
of 26 stars which showed these lines the best and at a good signal to noise
ratio. Such a selection enables us to measure the equivalent widths of these lines in a manner 
which is most reliable. The expected position of P18 (8437 \AA) is shown 
by a dashed line. As may be seen from the Figure 3, the Paschen line strengths
do show a monotonical increase with wavelength and then 
display a trend of flattening 
out around P17 and beyond. For our purpose of estimating the equivalent width of P18 
it looks reasonable from the figure to approximate the value by 
interpolating between the equivalent width values of P17 and P19, individually for each star. 
      
The flattening behavior of the Paschen line strengths beyond P17, 
shown in Figure 3, is possibly because these lines are optically thick and 
therefore being seen at a reduced intensity compared to 
that expected from recombination Case B analysis 
(\citet{Hummer87}; \citet{Storey95}). 
Similar behavior is seen for Brackett series lines \citep{steele01} and 
Humphrey and Pfund lines \citep{granada10} which is attributed to optical
depth effects. 

Subsequent to the correction of $W$(8446) from the contribution of P18, 
$W$(7774) and $W$(8446) are corrected for the underlying stellar
absorption. This is necessary because 
these lines come into emission only after filling the
underlying absorption component.
The strength of this absorption component is estimated from the synthetic
spectra of B0 ---B9 main sequence stars of solar metallicity given 
by \citet{munari05}
which are calculated from the SYNTHE code \citep{kurucz93} using NOVER
models as the input stellar atmospheres \citep{castelli97}.
We find $W$ values of the components for the B0 to B9 classes respectively
to be 0.006, 0.099, 0.143, 0.178, 0.209, 0.248, 0.272, 0.304, 0.345 and 0.400
\AA~respectively for the 7774 \AA~line and 0.003, 0.050, 0.074, 0.095, 0.111,
0.136, 0.153, 0.173, 0.199 and 0.236 for the 8446 \AA~line. These equivalent
width values of the 7774 and 8446 \AA~absorption components are added to their
corresponding emission values measured from the observed spectra to yield the
final $W$ values of the lines. The ratio of these equivalent widths is
converted into a flux ratio by multiplying
it with the ratio of continuum fluxes at 7774 and 8446 \AA~which is obtained
from \citet{kurucz92} models which are also given by Leitherer 
(\url{http://www.stsci.edu/science/starburst/Kurucz.html}). 
Typically the 7774/8446 continuum ratios from the
Kurucz models are in the range 1.34 to 1.26 for B0 to B9 and closely match the 
7774/8446 continuum ratio expected of a blackbody having an effective temperature of the
corresponding B type star's spectral class. 

The derived line flux ratio is finally corrected for extinction where the extinction
at 7774 and 8446 \AA~is derived  from the parameterized, seventh order polynomial fit
to the interstellar extinction given in \citet{cardelli89}. This gives $A(7774)$ = 0.6307$A_V$
and $A(8446)$ = 0.5275$A_V$ or equivalently
$A(7774)$ - $A(8446)$ =  0.32$\times$$E(B-V)$ with ratio of total-to-selective
extinction, R = 3.1. The $E(B-V)$ values of the CBe
stars are assumed to be equal to the mean $E(B-V)$
of the cluster with which it is associated. 
These values, available from the literature, are compiled in
\citet{mathew08} and were used for extinction correction and are shown in
column 9 of Table 1. However, it may be noted that the circumstellar envelope of a CBe star 
itself introduces an additional reddening expected to be of the order of
  $E(B-V)$ $\sim$ 0.1 magnitude (\citet{Dachs88}; \citet{slettebak85}). 
But this component is small compared to the cluster reddening 
(column 9, Table 1) and therefore should not significantly affect the dereddening corrections
to the O{\sc i} line flux ratio. 
The final  deredenned line flux ratio I(8446/7774), using the cluster $E(B-V)$
value, is given
in column 10 of Table 1 and plotted in the right hand panel of Figure 4. 
In case we include the 0.1 mag additional reddening due to the circumstellar
envelope then the I(8446/7774) ratio is found to change very marginally by 
$\sim$ 3 \%. 

\begin{figure}
\plotone{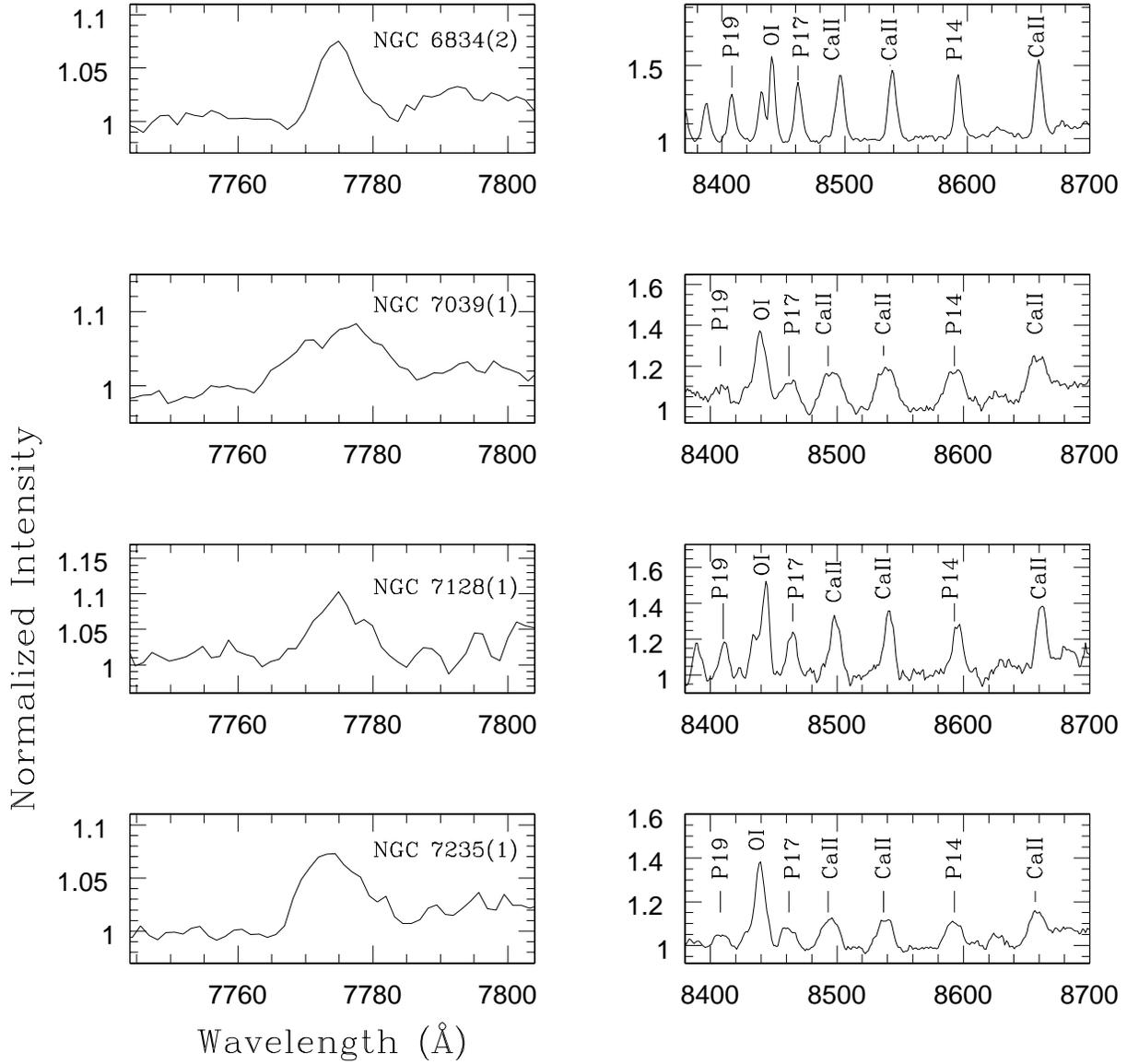}
\caption{O{\sc i} 7774 and 8446 line profiles of Group I candidates 
NGC 6834(2), NGC 7039(1), NGC 7128(1) and NGC 7235(1). 
Along with O{\sc i} 8446 which is blended with Hydrogen Paschen 18 (P18), other lines seen
are P14 8598 \AA~, P17 8467 \AA~, P19 8413 \AA~and Ca{\sc ii} 8498,
8542, 8662 \AA~. The nomenclature of these CBe stars is given in \citet{mathew11}}
\end{figure}

\begin{figure}
\plotone{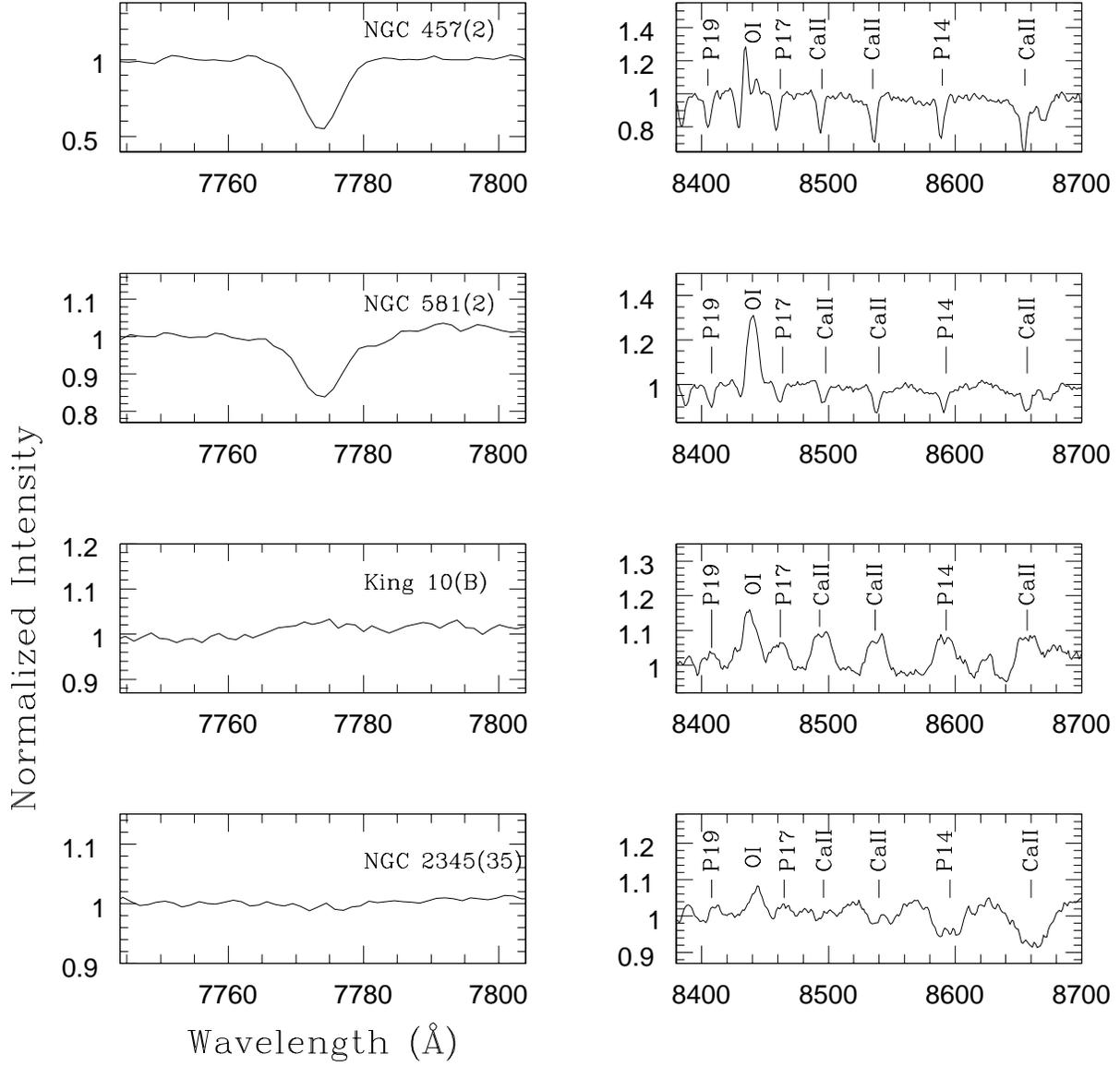}
\caption{O{\sc i} 7774 and 8446 line profiles of Group II candidates 
(1) NGC 457(2), (2) NGC 581(2) is shown in upper panel while that of Group III
  candidates (3) King 10(B), (4) NGC 2345(35) is shown in lower panel.}
\end{figure}

\begin{figure}
\plotone{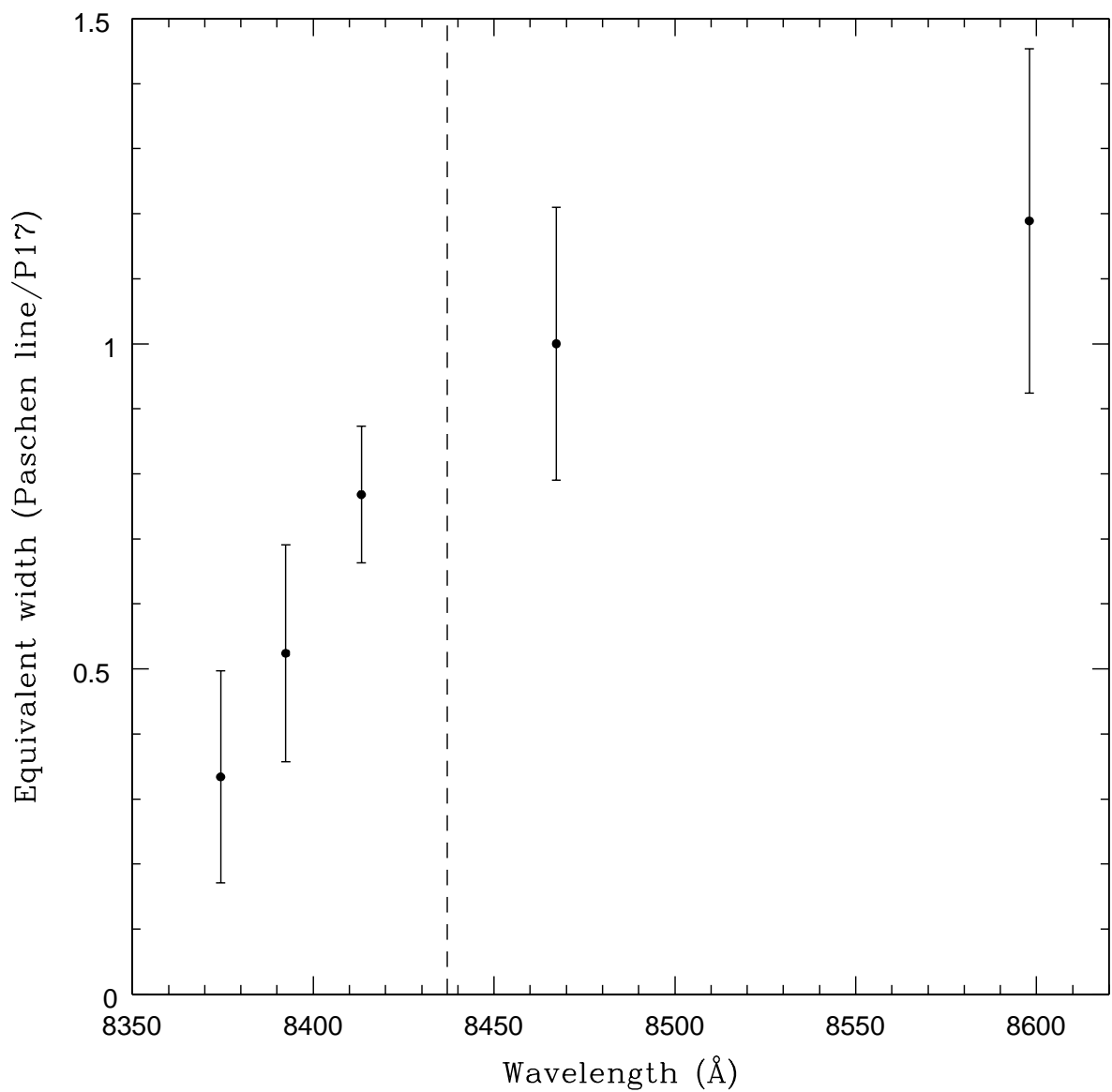}
\caption{The ratio of the mean equivalent widths of the unblended 
Paschen lines P14, P17, P19, P20 and  P21 normalized to the equivalent width
of P17 is shown in the figure. The corresponding wavelengths of these lines 
are 8598, 8467, 8413, 8392 and 8374 \AA~respectively. 
The wavelength position of P18 (8437 \AA) is shown in dashed line. 
Further details are given in the text in section 3.1.}
\end{figure}

\begin{figure}
\plotone{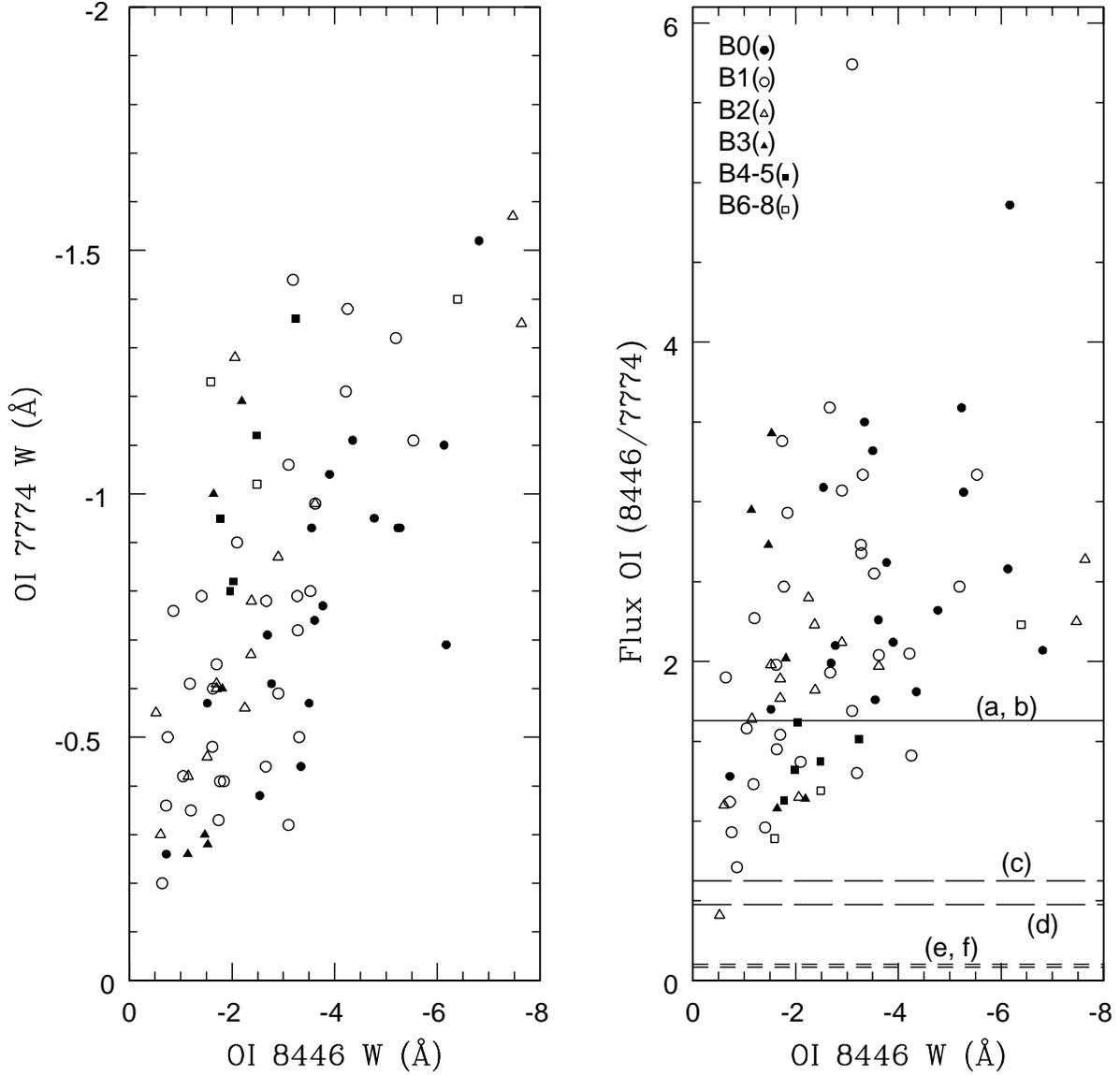}
\caption{Plot between equivalent widths of the 7774 and 8446 \AA~lines for the
  sample of CBe stars is shown in the left panel.
The flux ratio I(8446/7774) is shown in right panel. 
Horizontal lines indicate the expected values of
I(8446/7774), from KB95, in the case of pure collisional excitation. 
The values are (a) 1.62 for $T$ = 10000 K, $n_e$= 10$^{10}$ cm$^{-3}$, 
(b) 1.63 for $T$ = 20000 K, $n_e$= 10$^{10}$ cm$^{-3}$, 
(c) 0.626 for $T$ = 10000 K, $n_e$= 10$^{11}$ cm$^{-3}$, 
(d) 0.478 for $T$ = 20000 K, $n_e$= 10$^{11}$ cm$^{-3}$, 
(e) 0.102 for $T$ = 10000 K, $n_e$= 10$^{12}$ cm$^{-3}$ and 
(f) 0.083 for $T$ = 20000 K, $n_e$= 10$^{12}$ cm$^{-3}$. 
The stars are binned based on their spectral types; B0 -- filled circles, 
B1 -- open circles, B2 -- open triangles, B3 -- filled triangles, 
B4 - B5 -- filled squares and B6 - B8 -- open squares. }
\end{figure}

\subsection{Possibility of PAR process as the dominant excitation mechanism
  for O{\sc i} emission lines in CBe stars}
The expected I(8446/7774) ratio from KB95, in the absence of the PAR
process (i.e the parameter $R_p$ = -~$\infty$)  is shown in Figure 4
(right hand panel) in straight lines. Essentially $R_p$ = -~$\infty$
represents the case of pure collisional excitation case of BK95.
The I(8446/7774) ratio is shown at two temperatures of 10000 K and 20000 K
and three values of the electron density $n_e$= 10$^{10}$, 10$^{11}$ and
10$^{12}$ cm$^{-3}$. The choice of this parameter space for $n_e$ and $T$,
is a realistic approximation, of the actual range over which
these parameters vary in Be star disk.
There are several studies available in literature that model Be star disks in great detail
to estimate disk parameters 
(e.g. \citet{sigut07}; \citet{carciofi08}; \citet{gies07}). We find that a global view of
disk parameters is offered by the work of \citet{silaj10} who generate and compare model and
observed emission profiles for a statistically large
sample of 57 CBe stars. These models use the code of \citet{sigut07}.
The current models mentioned above assume that the
disk density decreases, starting from a base density value of $\rho_0$, with radial
distance $R$ as a power law with exponent $n$
(that is $\rho$ $\propto$ $\rho_0$$(R_{star}/R)^n$). \citet{silaj10} point out that
one of the interesting features of the derived disk density parameters is that a radial
power-law index of $n$ = 3.5 is strongly preferred for the model fits with
43\% of the fits, over all stars considered, requiring this power-law
index. We adopt this as a representative value for $n$. From their individually listed
values of $\rho_0$ for the entire sample, we find  a mean value of
1.09$\times$10$^{-10}$ gm cm$^{-3}$ for $\rho_0$ equivalent to a density of
6.5$\times$10$^{13}$ electrons cm$^{-3}$ assuming a completely ionized gas composed purely of
Hydrogen. With $n$ = 3.5 and a mean value of $\rho_0$ as above, and assuming a mean disk size of 6 to 8
$R_{star}$, the density ranges from 6.5$\times$10$^{13}$ at the inner edge to $\sim$ (0.5 to
1.2)$\times$10$^{11}$ electrons cm$^{-3}$ at the outer edge of the disk.
Thus a mean or representative value of $n_e$ across the disk should be taken as
1$\times$10$^{11}$ cm$^{-3}$  or greater; this is what we will adopt.
The above models also show that the disk is not isothermal but
the temperature of the matter in the disk is mostly seen to be in the range of
10000 to 20000 K.

Figure 4 (right panel) shows that for the adopted values of $n_e$ =
1$\times$10$^{11}$ cm$^{-3}$ and $T$ =10000 to 20000 K, the I(8446/7774)
ratio of all stars barring one, lies above the value predicted from pure
collisional excitation. This is also true for the case of $n_e$ = 10$^{12}$
cm$^{-3}$ and, from the trend, should be true for higher densities too.
The expected value of I(8446/7774) for $n_e$ = 10$^{10}$
cm$^{-3}$ is also shown,
more to illustrate the dependence on $n_e$ and $T$, but as argued before this
is likely a lower than actual value for the electron density.
Even in this case, a large number of stars show I(8446/7774) to be
significantly larger than expected. This shows that collisional excitation by
itself cannot account for the higher than expected I(8446/7774) ratio observed
in Be stars and the PAR process, which enhances the strength of O{\sc i}
$\lambda$8446, also contributes. In contrast, it may be mentioned that in
the solar chromosphere I(8446/7774) is found equal to 0.2 $\pm$ 0.02 by
\citet{penn99} who therefore concludes that  collisional excitation is more important
than the PAR process in O{\sc i} solar limb emission.

In Figure 4, the I(8446/7774) strengths at densities of $n_e$ = 10$^{10}$ and
10$^{12}$ are from KB95. But since they have listed their data for increments
in log $n_e$ in steps of 2 (i.e log $n_e$ = 4, 6, 8, 10 and 12), the value of
I(8446/7774) at other intermediate densities such as $n_e$ = 10$^{9}$, 10$^{11}$ etc were
obtained from the authors by writing to them (Bhatia, private communication).

Although the density exponent $n$ = 3.5 is favored by the \citet{silaj10} 
models, other values of the exponent could be considered too. But it may
be noted that the bulk of their profiles, 84 percent to be precise, are
modeled with $n$ $\le$ 3.5. If we thus consider lower values of $n$, the density
decline in the disk will be even slower in the disk compared to the $n$ = 3.5
case.  This will strengthen support for our arguments to consider a mean
density in the disk of 10$^{11}$ or more. An additional consideration may also
be taken into account. Though interferometric data is now gradually
emerging on emission zone size of lines like H$\alpha$ (e.g. \citet{Quirrenbach97}) and
a few other near-IR lines (e.g. \citet{gies07}, \citet{MillanGabet10}), no measurements are
yet reported for the size of the O{\sc i} line emitting region. But indirect
estimates about the sizes of line emitting regions have been reported by
\citet{Jaschek93} based on the observed velocity separation of the V
(violet) and R (red) components of Be line profiles and the model by
\citet{Huang72}. Based on the entire Jaschek data we have derived the O{\sc i} 8446
\AA~line emission region to have a mean size of 0.71 $\pm$ 0.27 of the
H$\alpha$ emission region size. 
This implies that O{\sc i} emission arises from relatively inner
regions of the disk and therefore regions of higher density. 
Following \citet{Jaschek93} we estimate the radius of the H$\alpha$ emission
region for the 10 CBe stars that showed double-peaked H$\alpha$ emission
line profiles from the sample studied by \citet{mathew11}. 
The mean value is around 12 R$_s$ and 3.5 R$_s$ for Keplerian and rigid body
rotation for the particles in the disk. This translates to 8.5 R$_s$ and 2.9 
R$_s$ for the size of O{\sc i} 8446 emission region since O{\sc i} 8446 line
emission region is 0.71 times the H$\alpha$ emission radius, as derived earlier.

In Figure 4 we have also indicated the spectral class of the stars. Visual
examination does not appear to indicate any significant correlation between the O{\sc i} ratio
and the spectral class. It may be expected that the temperature of the central
star, which is essentially a measure of the spectral class, could affect the
strength of the O{\sc i} 8446 line  via the amount of Lyman $\beta$ radiation
that the star emits. Eventhough this result looks promising, 
it needs to be strengthened further from the analysis of a large
sample of CBe stars, especially late-type CBe stars. 

If Ly$\beta$ fluorescence is operational in CBe stars one would 
expect a correlation between the emission strength of H$\alpha$ and O{\sc i} 
$\lambda$8446 lines, since H$\alpha$ and Ly$\beta$ both depend on the population of the
third hydrogen level \citep{bowen47}. 
\citet{Kitchin70} and \citet{Andrillat88} have studied the correlation
between the measured equivalent widths of H$\alpha$ and O{\sc i} $\lambda$8446
emission lines in CBe stars. With our simultaneous observations of O{\sc i} $\lambda$8446 and
H$\alpha$, it is interesting to derive a relation between their emission
line equivalent widths. The $W$(H$\alpha$) values for this analysis are for
the same sample of stars analysed here but whose equivalent width values are reported in 
\citet{mathew11} while $W$(8446) values are given in Table 1.  
From a simple linear fit, we have derived the following relation between 
$W$(H$\alpha$) and $W$(8446), both measured in \AA~, for Group I stars
as : 

\begin{equation}
W(8446) = 0.10 \times W(H\alpha) + 0.53 
\end{equation}

\begin{figure}
\plotone{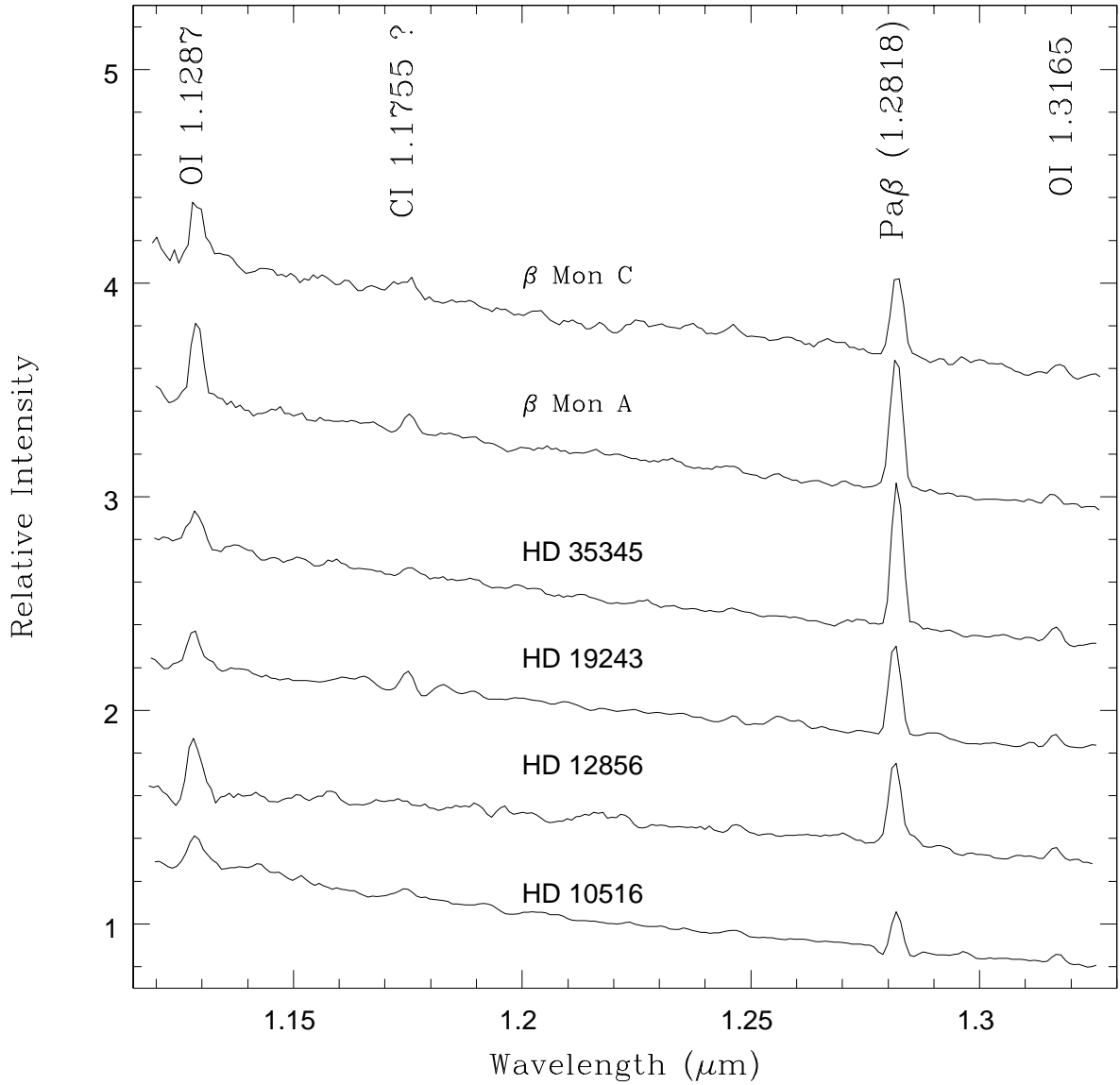}
\caption{The $J$ band spectra of selected CBe stars with the prominent lines
identified. The spectra of the stars HD 10516, HD 12856, HD 19243,
HD 35345, $\beta$ Mon A, $\beta$ Mon C are normalized with
respect to the band center at 1.22 $\mu$m.
There is offset between the adjacent spectra for clarity.}
\end{figure}

It is unlikely that recombination plays a dominant role in O{\sc i} excitation
since the I(8446/7774)
ratio, in the case of recombination, should be around 0.6 as indicated by
Grandi (1975, 1980) and \citet{strittmatter77}. However, this is not the case here with the 8446
\AA~line being stronger than the 7774 \AA~line in most cases contrary to expectations.
An interesting example where the I(8446/7774) ratio becomes large due to
the absence of recombination is in the Weigelt blobs around the LBV $\eta$
Carinae. As proposed by \citet{johansson05},
the front of the blobs are ionized providing the Lyman $\beta$ flux for the
PAR process while the inner and distal regions of the blobs remain neutral
because the ionizing flux cannot penetrate deep into the blobs. As a result
recombination does not occur while the PAR process dominates leading to the
extreme case of strong 8446 \AA~emission being seen while the 7774 \AA~line is completely absent.

Whether continuum fluorescence is the main source of excitation or not 
can be strongly constrained based on the ratio of strengths of the near-IR 11287 and 13165 \AA~lines.
It is expected that $W$(13165)/$W$(11287) $\ge$  1 if continuum fluorescence is the significant
excitation mechanism (\citet{strittmatter77}, \citet{grandi75}). On the other
hand, if the Ly$\beta$ fluorescence process dominates, the 11287 \AA~line should
become very strong as it is the primary line of the fluorescent cascade.
The $W$(13165)/$W$(11287) ratio is then expected to reverse and become smaller than unity.
In novae, a large number of which have been studied by us in the near-IR
(e.g. nova V1280 Sco -- \citet{das08}; nova V574 Puppis -- \citet{naik10}),
it is inevitably seen that $W$(11287) is several times larger than
$W$(13165) strongly supporting the dominance  of the PAR process in novae -
analysis of other optical O{\sc i} lines independently establishes this (KB
95). An extreme case of a nova with  extremely strong emission in the 11287
\AA~line accompanied by  no emission at 11365 \AA~is V4643 Sgr \citep{ashok06};
nova V1500 Cygni also shows similar and extremely strong fluorescent
lines of O{\sc i} \citep{strittmatter77}.

In Figure 5, we present the $J$ band spectra of selected CBe stars 
from a larger sample studied by us to
characterize their $J$ band behavior (paper in preparation). It is found that whenever
the 11287 and 13165 \AA~lines are present, as in the spectra shown, the former is always stronger of
the two. This is seen in the spectra of Figure 5 where a mean value of
$W$(13165)/$W$(11287) = 0.30 is found. Collisional excitation is also
predicted to give $W$(13165)/$W$(11287)  $\ge$  1 at $T$ = 10000 and
20000 K respectively for $n_e$ = 10$^{10}$ -- 10$^{12}$ cm$^{-3}$ (BK95). Thus both
continuum fluorescence and collisional excitation are expected to make the
13165 \AA~line stronger than 11287 \AA~line. This is not seen and 
the $W$(13165)/$W$(11287) ratio thus suggests that the Ly$\beta$ process 
has a significant role in exciting the near-IR lines in CBe stars. 
It may also be noted that continuum fluorescence if present, is expected to give
significant contribution to the O{\sc i} 7254 and 7002 \AA~lines \citep{strittmatter77}
and so also strong emission, comparable to O{\sc i} 8446 \AA~, in
the O{\sc i} 7990 \AA~line \citep{netzer76}.
No traces of these O{\sc i} lines are seen in our spectra within the detection
limits \citep{mathew11}. Based on the above evidence and arguments in favor of the PAR process
vis-a-vis other excitation mechanisms, we would conclude that the
Ly$\beta$ fluorescence is operational in CBe stars.

\acknowledgments
We thank the referee for his suggestions and comments which helped in improving the manuscript.
One of us (DPKB) thanks Dr. A. K. Bhatia for his help in providing additional
computational data to us. 
The research work at Physical Research Laboratory is funded by the Department
of Space, Government of India.

\newpage

\begin{deluxetable}{crrrrrrrrr}
\tabletypesize{\scriptsize}
\tablecaption{List of CBe stars which show O{\sc i} 7774 and 8446 in emission 
are shown with equivalent width and line flux ratio estimates. 
The coordinates and details of CBe stars are given in \citet{mathew11}\label{tbl-1}}
\tablewidth{0pt}
\tablehead{\colhead{Be star} & \colhead{Sp.type} & \colhead{Obs.date} &
  \colhead{$W$(7774)$^a$}& \colhead{$W$(8446)$^a$} & 
\colhead{$W$(7774)$^b$} & \colhead{$W$(8446)$^b$} &
\colhead{I(8446/7774)} & \colhead{$E(B-V)$} & \colhead{I$_C$$^c$} }
\startdata
Berkeley86(26)   &  B1V     & 27-06-2005 & -0.31   &  -1.72   &  -0.41   &  -1.77    &   3.22    &     0.90   &   2.47 \\
Berkeley87(3)    &  B2      & 08-10-2005 & -0.41   &  -0.45   &  -0.55   &  -0.52    &   0.71    &     1.90   &   0.41 \\
Berkeley87(4)    &  B0-1V   & 09-10-2005 & -0.45   &  -3.29   &  -0.50   &  -3.31    &   4.98    &     1.53   &   3.17 \\
Berkeley90(1)    &  B0V     & 28-08-2006 & -0.76   &  -3.77   &  -0.77   &  -3.77    &   3.68    &     1.15   &   2.62 \\
Collinder96(1)   &  B0-1V   & 21-11-2005 & -0.36   &  -1.81   &  -0.41   &  -1.84    &   3.37    &     0.48   &   2.93 \\
IC4996(1)        &  B3      & 15-07-2005 & -0.10   &  -1.44   &  -0.28   &  -1.53    &   4.23    &     0.71   &   3.43 \\
King10(A)        &  B1V     & 29-07-2005 & -0.50   &  -1.58   &  -0.60   &  -1.63    &   2.03    &     1.13   &   1.45 \\
King10(C)        &  B2V     & 30-07-2005 & -0.16   &  -0.54   &  -0.30   &  -0.61    &   1.54    &     1.13   &   1.10 \\
King10(E)        &  B3V     & 31-07-2005 & -0.12   &  -1.38   &  -0.30   &  -1.47    &   3.81    &     1.13   &   2.73 \\
NGC436(2)        &  B5-7V   & 10-01-2007 & -0.96   &  -1.44   &  -1.23   &  -1.59    &   1.03    &     0.50   &   0.89 \\
NGC436(5)        &  B3V     & 09-01-2007 & -0.42   &  -1.72   &  -0.60   &  -1.81    &   2.34    &     0.50   &   2.02 \\
NGC457(1)        &  B3V     & 29-09-2006 & -0.08   &  -1.05   &  -0.26   &  -1.14    &   3.41    &     0.49   &   2.95 \\
NGC581(1)        &  B2V     & 28-09-2006 & -0.46   &  -1.63   &  -0.60   &  -1.70    &   2.15    &     0.44   &   1.89 \\
NGC581(3)        &  B0-1V   & 28-09-2006 & -0.30   &  -1.18   &  -0.35   &  -1.20    &   2.58    &     0.44   &   2.27 \\
NGC654(2)        &  B0-1    & 29-09-2006 & -1.01   &  -3.08   &  -1.06   &  -3.10    &   2.20    &     0.90   &   1.69 \\
NGC659(1)        &  B2V     & 21-11-2005 & -0.53   &  -2.30   &  -0.67   &  -2.37    &   2.68    &     0.63   &   2.23 \\
NGC659(3)        &  B1V     & 21-11-2005 & -0.40   &  -0.70   &  -0.50   &  -0.75    &   1.12    &     0.63   &   0.93 \\
NGC663(1)        &  B5V     & 08-10-2005 & -0.70   &  -1.64   &  -0.95   &  -1.78    &   1.43    &     0.80   &   1.13 \\
NGC663(2)        &  B0-1V   & 07-10-2005 & -0.39   &  -2.63   &  -0.44   &  -2.66    &   4.54    &     0.80   &   3.59 \\
NGC663(4)        &  B1V     & 24-10-2005 & -0.38   &  -1.57   &  -0.48   &  -1.62    &   2.51    &     0.80   &   1.98 \\
NGC663(5)        &  B1V     & 22-11-2005 & -1.34   &  -3.14   &  -1.44   &  -3.19    &   1.65    &     0.80   &   1.30 \\
NGC663(9)        &  B1V     & 24-10-2005 & -1.11   &  -4.17   &  -1.21   &  -4.22    &   2.60    &     0.80   &   2.05 \\
NGC663(11)       &  B2V     & 09-10-2005 & -0.42   &  -2.18   &  -0.56   &  -2.25    &   3.04    &     0.80   &   2.40 \\
NGC663(12V)      &  B0-1V   & 21-11-2005 & -0.27   &  -3.07   &  -0.32   &  -3.10    &   7.27    &     0.80   &   5.74 \\
NGC663(15)       &  B1V     & 25-10-2005 & -0.80   &  -2.05   &  -0.90   &  -2.10    &   1.74    &     0.80   &   1.37 \\
NGC663(16)       &  B1V     & 25-10-2005 & -0.55   &  -1.65   &  -0.65   &  -1.70    &   1.95    &     0.80   &   1.54 \\
NGC663(P5)       &  B2V     & 14-10-2005 & -0.64   &  -2.31   &  -0.78   &  -2.38    &   2.31    &     0.80   &   1.82 \\
NGC663(P8)       &  B2V     & 14-10-2005 & -0.32   &  -1.45   &  -0.46   &  -1.52    &   2.51    &     0.80   &   1.98 \\
NGC663(P23)      &  B2V     & 25-10-2005 & -0.28   &  -1.08   &  -0.42   &  -1.15    &   2.08    &     0.80   &   1.64 \\
NGC663(P25)      &  B0-1V   & 22-11-2005 & -0.15   &  -0.61   &  -0.20   &  -0.64    &   2.41    &     0.80   &   1.90 \\
NGC869(1)        &  B0V     & 21-01-2006 & -0.92   &  -5.23   &  -0.93   &  -5.23    &   4.23    &     0.56   &   3.59 \\
NGC869(2)        &  B0-1V   & 20-01-2006 & -0.56   &  -1.16   &  -0.61   &  -1.18    &   1.45    &     0.56   &   1.23 \\
NGC869(5)        &  B1V     & 20-01-2006 & -0.32   &  -1.00   &  -0.42   &  -1.05    &   1.86    &     0.56   &   1.58 \\
NGC884(1)        &  B0-1V   & 22-01-2006 & -0.71   &  -0.83   &  -0.76   &  -0.86    &   0.84    &     0.56   &   0.71 \\
NGC884(2)        &  B0-1V   & 22-01-2006 & -1.06   &  -5.50   &  -1.11   &  -5.53    &   3.74    &     0.56   &   3.17 \\
NGC884(5)        &  B0V     & 28-09-2006 & -0.56   &  -1.52   &  -0.57   &  -1.52    &   2.00    &     0.56   &   1.70 \\
NGC957(1)        &  B0-1V   & 07-12-2005 & -0.67   &  -3.26   &  -0.72   &  -3.28    &   3.43    &     0.84   &   2.68 \\
NGC1220(1)       &  B5V     & 21-11-2005 & -0.87   &  -2.34   &  -1.12   &  -2.48    &   1.68    &     0.70   &   1.37 \\
NGC1893(1)       &  B1V     & 21-11-2005 & -1.22   &  -5.14   &  -1.32   &  -5.19    &   2.93    &     0.58   &   2.47 \\
NGC2345(27)      &  B3V     & 28-12-2007 & -1.01   &  -2.10   &  -1.19   &  -2.19    &   1.43    &     0.77   &   1.14 \\
NGC2345(59)      &  B3V     & 07-12-2005 & -0.82   &  -1.55   &  -1.00   &  -1.64    &   1.27    &     0.54   &   1.08 \\
NGC2345(X2)      &  B5V     & 15-12-2007 & -0.57   &  -1.89   &  -0.82   &  -2.03    &   1.88    &     0.50   &   1.62 \\
NGC2414(2)       &  B1V     & 07-12-2005 & -0.23   &  -1.69   &  -0.33   &  -1.74    &   3.93    &     0.51   &   3.38 \\
NGC2421(1)       &  B1V     & 21-01-2006 & -0.69   &  -3.22   &  -0.79   &  -3.27    &   3.09    &     0.42   &   2.73 \\
NGC6649(1)       &  B0V     & 09-06-2007 & -0.70   &  -2.69   &  -0.71   &  -2.69    &   2.84    &     1.20   &   1.99 \\
NGC6649(6)       &  B5V     & 10-06-2007 & -0.55   &  -1.84   &  -0.80   &  -1.97    &   1.88    &     1.20   &   1.32 \\
NGC6649(7)       &  B2V     & 10-06-2007 & -0.84   &  -3.55   &  -0.98   &  -3.62    &   2.80    &     1.20   &   1.97 \\
NGC6834(1)       &  B5V     & 07-10-2005 & -1.11   &  -3.10   &  -1.36   &  -3.23    &   1.81    &     0.61   &   1.51 \\
NGC6834(2)       &  B1V     & 07-10-2005 & -0.49   &  -2.85   &  -0.59   &  -2.90    &   3.67    &     0.61   &   3.07 \\
NGC7039(1)       &  B1-3V   & 21-11-2005 & -1.14   &  -1.99   &  -1.28   &  -2.06    &   1.22    &     0.19   &   1.15 \\
NGC7128(1)       &  B1V     & 14-10-2005 & -0.88   &  -3.57   &  -0.98   &  -3.62    &   2.76    &     1.03   &   2.04 \\
NGC7235(1)       &  B0-1V   & 14-10-2005 & -0.75   &  -3.51   &  -0.80   &  -3.53    &   3.32    &     0.90   &   2.55 \\
NGC7261(1)       &  B0V     & 07-12-2005 & -1.03   &  -3.90   &  -1.04   &  -3.90    &   2.82    &     0.97   &   2.12 \\
NGC7261(2)       &  B1V     & 07-12-2005 & -0.26   &  -0.67   &  -0.36   &  -0.72    &   1.49    &     0.97   &   1.12 \\
NGC7261(3)       &  B0-1V   & 07-12-2005 & -0.73   &  -2.65   &  -0.78   &  -2.67    &   2.57    &     0.97   &   1.93 \\
NGC7380(2)       &  B1-3V   & 17-07-2006 & -0.73   &  -2.83   &  -0.87   &  -2.90    &   2.53    &     0.60   &   2.12 \\
NGC7380(3)       &  B1-3V   & 17-07-2006 & -0.47   &  -1.63   &  -0.61   &  -1.70    &   2.11    &     0.60   &   1.77 \\
NGC7419(A)  &  B0V   &  15-07-2005 & -0.43 &  -3.34 &  -0.44 &  -3.34 &    5.70  &         1.65 &   3.50 \\
NGC7419(B)  &  B0V   &  27-06-2005 & -0.92 &  -3.55 &  -0.93 &  -3.55 &    2.87  &         1.65 &   1.76 \\
NGC7419(D)  &  B1V   &  15-07-2005 & -1.28 &  -4.20 &  -1.38 &  -4.25 &    2.30  &         1.65 &   1.41 \\
NGC7419(E)  &  B6    &  31-07-2005 & -1.13 &  -6.24 &  -1.40 &  -6.39 &    3.62  &         1.65 &   2.23 \\
NGC7419(G)  &  B0V   &  08-08-2005 & -1.10 &  -4.35 &  -1.11 &  -4.35 &    2.95  &         1.65 &   1.81 \\
NGC7419(I)  &  B0V   &  15-07-2005 & -0.60 &  -2.77 &  -0.61 &  -2.77 &    3.41  &         1.65 &   2.10 \\
NGC7419(J)  &  B0V   &  08-08-2005 & -0.25 &  -0.72 &  -0.26 &  -0.72 &    2.08  &         1.65 &   1.28 \\
NGC7419(L)  &  B0V   &  15-07-2005 & -0.73 &  -3.61 &  -0.74 &  -3.61 &    3.67  &         1.65 &   2.26 \\
NGC7419(M)  &  B0V   &  08-08-2005 & -0.37 &  -2.54 &  -0.38 &  -2.54 &    5.02  &         1.65 &   3.09 \\
NGC7419(N)  &  B2.5  &  08-08-2005 & -1.41 &  -7.39 &  -1.57 &  -7.47 &    3.66  &         1.65 &   2.25 \\
NGC7419(O)  &  B8V   &  08-08-2005 & -0.68 &  -2.29 &  -1.02 &  -2.49 &    1.94  &         1.65 &   1.19 \\
NGC7419(Q)  &  B1-3V &  08-08-2005 & -1.21 &  -7.57 &  -1.35 &  -7.64 &    4.29  &         1.65 &   2.64 \\
NGC7419(1)  &  B0V   &  07-10-2005 & -0.94 &  -4.77 &  -0.95 &  -4.77 &    3.77  &         1.65 &   2.32 \\
NGC7419(3)  &  B0V   &  08-08-2005 & -1.51 &  -6.81 &  -1.52 &  -6.81 &    3.37  &         1.65 &   2.07 \\
NGC7419(4)  &  B0V   &  07-10-2005 & -0.16 &  -2.39 &  -0.17 &  -2.39 &    10.6  &         1.65 &   6.52 \\
NGC7419(5)  &  B0V   &  07-10-2005 & -1.09 &  -6.13 &  -1.10 &  -6.13 &    4.19  &         1.65 &   2.58 \\
NGC7510(1A) &  B0V   &  24-10-2005 & -0.56 &  -3.50 &  -0.57 &  -3.50 &    4.62  &         1.12 &   3.32 \\
NGC7510(1B) &  B1V   &  12-10-2005 & -0.69 &  -1.36 &  -0.79 &  -1.41 &    1.33  &         1.12 &   0.96 \\
NGC7510(1C) &  B0V   &  13-10-2005 & -0.92 &  -5.27 &  -0.93 &  -5.27 &    4.26  &         1.12 &   3.06 \\
Roslund4(2) &  B0V   &  25-10-2005 & -0.68 &  -6.17 &  -0.69 &  -6.17 &    6.72  &         1.10 &   4.86 \\
\enddata
\tablenotetext{a}{Observed equivalent widths in \AA.}
\tablenotetext{b}{Equivalent widths in \AA~corrected for underlying stellar
  absorption component.}
\tablenotetext{c}{This column gives the extiction corrected line flux ratio 
of I(8446/7774).}
\end{deluxetable}


\begin{thebibliography}{}
\bibitem[Andrillat \& Fehrenbach(1982)]{Andrillat82} Andrillat, Y., Fehrenbach, Ch. 1982, \aaps, 48, 93
\bibitem[Andrillat, Jaschek \& Jaschek(1988)]{Andrillat88} Andrillat, Y., Jaschek, M., Jaschek, C. 1988, \aaps, 72, 129
\bibitem[Ashok et al.(2006)]{ashok06} Ashok, N. M., Banerjee, D. P. K., Varricatt, W. P., Kamath, U. S. 2006, \mnras, 368, 592
\bibitem[Banerjee, Rawat \& Janardhan(2000)]{Banerjee00} Banerjee, D. P. K.,
  Rawat, S. D., Janardhan, P. 2000, \aaps, 147, 229
\bibitem[Bhatia \& Kastner(1995)]{bhatia95} Bhatia, A. K., Kastner, S. O. 1995, \apjs, 96, 325
\bibitem[Bowen(1947)]{bowen47} Bowen, I. S. 1947, \pasp, 59, 196
\bibitem[Briot(1981a)]{briot81a} Briot, D. 1981a, \aap, 103, 1
\bibitem[Briot(1981b)]{briot81b} Briot, D. 1981b, \aap, 103, 5
\bibitem[Burbidge(1952)]{burbidge52} Burbidge, E. M. 1952, \apj, 115, 418
\bibitem[Carciofi \& Bjorkman(2008)]{carciofi08} Carciofi, A. C., Bjorkman, J. E. 2008, \apj, 684, 1374
\bibitem[Cardelli, Clayton \& Mathis(1989)]{cardelli89} Cardelli, J. A., Clayton, G. C., Mathis, J. S. 1989, \apj, 345, 245
\bibitem[Castelli, Gratton \& Kurucz(1997)]{castelli97} Castelli, F., Gratton, R. G., Kurucz, R. L. 1997, \aap, 318, 841
\bibitem[Clark \& Steele(2000)]{clark00} Clark, J. S., Steele, I. A. 2000, \aaps, 141, 65
\bibitem[Collins(1987)]{collins87} Collins, G. W. 1987, in IAU Colloq. 92, Physics of Be Stars, ed. A. Slettebak \& T. P. Snow (Cambridge: Cambridge Univ. Press), 3
\bibitem[Dachs et al.(1986)]{Dachs86} Dachs, J., Hanuschik, R., Kaiser, D.,
  Ballereau, D., Bouchet, P. 1986, \aaps, 63, 87
\bibitem[Dachs, Kiehling \& Engels(1988)]{Dachs88} Dachs, J., Kiehling, R., Engels, D. 1988, \aap, 194, 167
\bibitem[Dachs, Hummel \& Hanuschik(1992)]{Dachs92} Dachs, J., Hummel, W., Hanuschik, R. W. 1992, \aaps, 95, 437
\bibitem[Das et al.(2008)]{das08} Das, R. K., Banerjee, D. P. K., Ashok, N. M., Chesneau, O. 2008, \mnras, 391, 1874
\bibitem[Gies et al.(2007)]{gies07} Gies, D. R. et al. 2007, \apj, 654, 527
\bibitem[Granada, Arias \& Cidale(2010)]{granada10} Granada, A., Arias, M. L., Cidale, L. S. 2010, \aj, 139, 1983
\bibitem[Grandi(1975)]{grandi75} Grandi, S. A. 1975, \apj, 196, 465
\bibitem[Grandi(1980)]{grandi80} Grandi, S. A. 1980, \apj, 238, 10
\bibitem[Hanuschik(1986)]{hanuschik86} Hanuschik, R. W. 1986, \aap, 166, 185 
\bibitem[Hanuschik(1987)]{hanuschik87} Hanuschik, R. W.	1987, \aap, 173, 299
\bibitem[Huang(1972)]{Huang72} Huang, Su-Shu. 1972, \apj, 171, 549
\bibitem[Hummer \& Storey(1987)]{Hummer87} Hummer, D. G., Storey, P. J.	1987, \mnras, 224, 801
\bibitem[Jaschek \& Jaschek(1993)]{Jaschek93} Jaschek, C., Jaschek, M. 1993, \aaps, 97, 807
\bibitem[Johansson \& Letokhov(2005)]{johansson05} Johansson, S., Letokhov, V. S. 2005, \mnras, 364, 731
\bibitem[Kastner \& Bhatia(1995)]{kastner95} Kastner, S. O., Bhatia, A. K. 1995, \apj, 439, 346
\bibitem[Kitchin \& Meadows(1970)]{Kitchin70} Kitchin, C. R., Meadows, A. J. 1970, \apss, 8, 463
\bibitem[Kurucz(1992)]{kurucz92} Kurucz, R. L., 1992, in IAU Symposium 149, The Stellar Populations
  of Galaxies, ed. B. Barbuy \& A. Renzini (Dordrecht:Kluwer), 225
\bibitem[Kurucz(1993)]{kurucz93} Kurucz, R., 1993, SYNTHE Spectrum Synthesis Programs and Line Data, CD-ROM No. 18
\bibitem[Mathew, Subramaniam \& Bhatt(2008)]{mathew08} Mathew, B., Subramaniam, A., Bhatt, B. C. 2008, \mnras, 388, 1879
\bibitem[Mathew \& Subramaniam(2011)]{mathew11} Mathew, B., Subramaniam, A., 2011, BASI, 39, 517
\bibitem[Meilland et al.(2012)]{meilland12} Meilland, A., Millour, F., Kanaan,
  S., Stee, Ph., Petrov, R., Hofmann, K.-H., Natta, A., Perraut, K. 2012, \aap, 538A, 110
\bibitem[Millan-Gabet et al.(2010)]{MillanGabet10} Millan-Gabet, R. et al. 2010, \apj, 723, 544
\bibitem[Munari et al.(2005)]{munari05} Munari U., Sordo R., Castelli F., Zwitter T. 2005, \aap, 442, 1127
\bibitem[Naik et al.(2010)]{naik10} Naik, S., Banerjee, D. P. K., Ashok, N. M., Das, R. K. 2010, \mnras, 404, 367
\bibitem[Netzer \& Penston(1976)]{netzer76} Netzer, H., Penston, M. V. 1976, \mnras, 174, 319
\bibitem[Penn(1999)]{penn99} Penn, M. J. 1999, \apj, 518L, 131P
\bibitem[Porter \& Rivinius(2003)]{porter03} Porter, J. M., Rivinius, T. 2003, \pasp, 115, 1153
\bibitem[Quirrenbach et al.(1997)]{Quirrenbach97} Quirrenbach, A. et al. 1997, \apj, 479, 477
\bibitem[Sigut \& Jones(2007)]{sigut07} Sigut, T. A. A., Jones, C. E., 2007, \apj, 668, 481
\bibitem[Silaj et al.(2010)]{silaj10} Silaj, J., Jones, C. E., Tycner, C., Sigut, T. A. A., Smith, A. D. 2010, \apjs, 187, 228
\bibitem[Slettebak(1951)]{slettebak51} Slettebak, A. 1951, \apj, 113, 436
\bibitem[Slettebak(1982)]{slettebak82} Slettebak, A. 1982, \apjs, 50, 55
\bibitem[Slettebak(1985)]{slettebak85} Slettebak, A. 1985, \apjs, 59, 769
\bibitem[Steele \& Clark(2001)]{steele01} Steele, I. A., Clark, J. S. 2001, \aap, 371, 643
\bibitem[Storey \& Hummer(1995)]{Storey95} Storey, P. J., Hummer, D. G. 1995, \mnras, 272, 41
\bibitem[Strittmatter et al.(1977)]{strittmatter77} Strittmatter, P. A. et al. 1977, \apj, 216, 23
\end{thebibliography}
\end{document}